\documentclass{article}
\usepackage{authblk}
\usepackage{graphicx} 

\usepackage{amsfonts}
\usepackage{amsmath}
\usepackage{amsthm}
\usepackage{amssymb}
\usepackage{verbatim}
\usepackage{xcolor}

\usepackage{graphicx} 
\usepackage{overpic}
\usepackage[final]{hyperref}

\newcommand{\comm}[1]{}

\def\ti{\tilde}

\def\({\left(}
\def\){\right)}
\def\oli{\overline}

\def\raw{\rightarrow}

\def\no={\neq}
\def\sm{\setminus}

\def\R{{\mathbb R}}

\def\DD{{\mathcal D}}

\def\SS{{\mathcal S}}

\def\al{\alpha}

\def\vep{\varepsilon}

\def\la{\lambda}

\def\Om{\Omega}

\newtheorem{Thm}{Theorem}[section]
\newtheorem{Defn}[Thm]{Definition}
\newtheorem{Prop}[Thm]{Proposition}

\newtheorem{Sublem}{SubLemma}

\newtheorem{Def}[Thm]{Definition}

\usepackage[backend=biber, style=authoryear, natbib=true]{biblatex}
\usepackage{graphicx}
\graphicspath{ {./images/} }

\title{Necessary conditions for stable equilibria in the Lotka-Volterra equations}
\author[1]{M. Aspenberg}
\author[1,2]{ E. A. Martens}
\author[3,4]{K. Wollein Waldetoft}
\date{}

\affil[1]{Centre for Mathematical Sciences, Lund University, Lund, Sweden.}
\affil[2]{Centre for Mathematical Modeling - Health and Disease, Institute for Mathematics and Physics, Dept. of Environment and Science, Roskilde University, Denmark}
\affil[3]{Department of Infectious Diseases, Uppsala University Hospital, Uppsala, Sweden}
\affil[4]{Faculty of Medicine, Department of Clinical Sciences, Division of Infection Medicine, Lund University, Lund, Sweden}

\begin{document}

\maketitle


\begin{abstract}
We consider the Lotka-Volterra system and  provide necessary conditions for an equilibrium to be stable. Our result naturally complements earlier fundamental results by N. Adachi, Y. Takeuchi, and H. Tokumaru, and B.S. Goh who gave sufficient conditions for the existence of a stable equilibrium point. 
\end{abstract}

\noindent
 {\bf Acknowledgements:} The first author gratefully acknowledges support from the Swedish Research Council. 
\vspace{3mm}


\section{Introduction}

In population dynamics, a classical model is the Lotka-Volterra equations (LV) (\cite{Volterra-nat, Volterra-fluct, Lotka-book}), 
\begin{equation} \label{LV}
   x_i'(t)= r_i x_i(t) \left(1 - \frac{1}{K_i} \sum_{j =1}^n \al_{ij} x_j(t) \right),
\end{equation}
where $i = 1, \ldots , n$ and $x_i$ are the abundances of $n$ given species. The $K_i >0$ are the carrying capacities and $r_i > 0$ the intrinsic growth rates. The numbers $\al_{ij}$, $i,j=1, \ldots, n$ form an $n \times n$-matrix $A$ with elements $(A)_{ij} = \al_{ij}$. Each number $\al_{ij}$ reflects the interaction between species $x_j$ and $x_i$. More precisely, $\al_{ij}$ is a measure of how $x_j$ influences $x_i$; if $\al_{ij} < 0$, $x_j$ facilitates $x_i$, and if $\al_{ij} > 0$, $x_j$ inhibits $x_i$. A higher value of $|\al_{ij}|$ indicates that the interaction is stronger. The matrix does not have to be symmetric. For example, if $\al_{ij} < 0$ and $\al_{ji} > 0$, then $x_j$ is facilitatory towards $x_i$, but $x_i$ is inhibitory towards $x_j$. However, the self interactions are usually set to $\al_{ii} = 1$, for all $i=1,\ldots,n$, and we also assume this in our study. 

Our study focuses on finding conditions for the existence of stable equilibrium points in the  LV-equations (\ref{LV}). This problem turns out to be both an interesting and complicated, in particular for large systems where $n$ is large,  and goes back to a series of studies (\cite{May-book,MacArthur, Go-Ma-Mo, Goh, Cronin}) and (\cite{Ad-Ta-To-1,Ad-Ta-To-2, Ad-Ta-1, Ad-Ta-2, Ad-Ta-3}) \textit{et al.} 

The LV-equations (\ref{LV}) find many applications in biology and are also very interesting from a mathematical point of view. It turns out that the system exhibits  already for low dimensions rich dynamics (\cite{Smale-examples}) including  chaotic behaviour (\cite{Arn-Cou-Pey-Tre, Arn-Cou-Tre, Vano-etal}). For a thorough study of completely competitive or completely collaborative systems see e.g. (\cite{Hirsch-II, Hirsch-III}). 

We begin by reviewing some fundamental results (\cite{Ad-Ta-To-1,Ad-Ta-To-2, Ad-Ta-1, Ad-Ta-2, Ad-Ta-3}) and (\cite{Goh}) on the existence of a stable equilibrium point. These results usually (but not always) state sufficient conditions for existence and stability. We will then turn to the result of the present paper, stating necessary conditions for stability. 

\section{Preliminaries and earlier results}

An equilibrium point for the Lotka-Volterra equations (LV) satisfies, by definition, 
\[
\oli{x}'(t) = (x_1'(t), \ldots, x_n'(t)) = (0,\ldots, 0). 
\]
Clearly, an equilibrium point is not dependent on time, and we denote it by $\oli{x}^* = (x_1^*, \ldots, x_n^*)$. 
We are only interested in solutions in the first orthant $\R_+^n = \{ (x_1, \ldots, x_n) : x_j \geq 0 \text{ all } 1 \leq i \leq n \}$. Equilibrium points in the first orthant are called {\em admissible}. When we talk about equilibria, in general, we implicitly assume that they are admissible. So for each $1 \leq i \leq n$, at least one of the following two equations must hold:
\begin{align}
    x_i &= 0 \\
    \al_{i1} x_1 + \al_{i2} x_2 + \ldots + \al_{in} x_n &= K_i. 
\end{align}

Of course there may be several equilibria in the first orthant. If there is an equilibrium $\oli{x}^*$ for which all $x_i^* > 0$, then we call it {\em feasible}. These points are of special interest. If we have an equilibrium $\oli{x}^*$ for which $x_i = 0$ for indices $i$ belonging to some index set $I \subset \{ 1,\ldots,n \}$, and $x_i > 0$ for $i \notin I$, then we say that it is {\em sub-feasible} if $0 < |I| < n$, and {\em non-feasible} if $|I|=n$. The number $|I|$ is called the {\em order} of the equilibrium. Hence a feasible equilibrium has order $0$ (it can be viewed as a sub-feasible equilibrium of order $0$). The number of surviving species in an LV-system depends heavily on the interaction coefficients, see (\cite{Clenet-Massol-Najim}) for results on the distribution of surviving species when $\al_{ij}$ are random. 

If we look for feasible equilibria, we consider the nullclines 
\[
P_i: \al_{i1} x_1 + \al_{i2} x_2 + \ldots + \al_{in} x_n = K_i. 
\]
We will assume that the normal vectors of the hyperplanes are linearly independent, and moreover, that these normal vectors together with the unit vectors in $\R^n$ are linearly independent. This implies, in particular, that the nullclines intersect at a unique point. Obviously, if this point belongs to the first orthant, it is an admissible equilibrium, and we put
\[
\oli{x}^*  = \bigcap_{j=1}^n P_j.
\]
Again, if all $x_i^* > 0$ it is feasible. In this case we solve $A\oli{x} = \oli{K} = (K_1, K_2, \ldots, K_n)$, and assume that $A$ is invertible,
\[
\oli{x}^* = A^{-1} \oli{K}.
\]


A principal sub-matrix of an $n \times n$-matrix $A$ is a matrix where we have deleted a set $S \subset \{1, \ldots, n \}$ of rows and columns (the same set for both the rows and columns) from $A$. In the following, we denote by $\DD_k(A)$ all principal sub-matrices of $A$ of size $k$.

Following (\cite{Ad-Ta-To-1, Ad-Ta-To-2}), the matrix $A$ belongs to the class $\SS$ if there exists a diagonal matrix $W$ such that
\[
WA + A^TW
\]
is positive definite. We have the following fundamental result from (\cite{Ad-Ta-To-1}) and (\cite{Goh}): 
\begin{Thm}
    If the matrix $A$ belongs to $\SS$ then there is a unique stable equilibrium for the corresponding Lotka-Volterra system. 
\end{Thm}

Recall that a minor of $A$ is a determinant of a principal sub-matrix of $A$. 
\begin{Def}
The matrix $A$ is a $P$-matrix if all minors of $A$ are positive.      
\end{Def}
In (\cite{Ad-Ta-To-1}), it is also proven that:
\begin{Thm}
    If $A$ belongs to $\SS$, then $A$ is a $P$-matrix. 
\end{Thm}
If all off-diagonal elements are non-positive, the same authors prove both sufficient and necessary conditions for stability, namely:
\begin{Thm}
Suppose that all off-diagonal elements of $A$ are non-positive, i.e., $\al_{ij} \leq 0$ whenever $i \neq j$. Then, if there is an equilibrium point $\oli{x}^* \in \R_+^n$, it is stable if and only if $A$ is a $P$-matrix. 
\end{Thm}

For further results on the stability and existence of (stable) equilibrium points, see also ~(\cite{Ad-Ta-To-2, Ad-Ta-1, Ad-Ta-2, Ad-Ta-3}).  

\section{Necessary conditions for stability}

In this section we present the main result of the present study. 
\begin{Defn}
    An equilibrium point,
    $\oli{x}^* = (x_1^*, \ldots, x_n^*)$,  is \emph{feasible} if $x_i^* > 0$ for all $i\in[n]:=\{1,\ldots,n\}$; if, on the other hand,  $x_i^* = 0$ for $i\in I\subseteq [n]$ and $x_i^*>0$ for $i\in [n]\setminus I$, it is \emph{sub-feasible} of order $|I|$.
\end{Defn}
Let $D^*$ be the diagonal matrix with elements $(r_1/K_1)x_1^*, \ldots,(r_n/K_n)x_n^*$ along the diagonal (from the top left to the bottom right). 
We get the following necessary condition for stability. 
\begin{Thm} \label{equi-s}
Let $B = D^* A$. Suppose there is a feasible equilibrium. Then, if it is stable, we must have
   \begin{equation} \label{sum-1}
   \sum_{C \in \DD_k(B)} \det(C) > 0
   \end{equation}
for all $k=0, \ldots n$.

For a sub-feasible equilibrium $\oli{x}^*$ of order $k$, suppose that $x_i^* = 0$ for $i \in I$. Then a necessary condition for stability is that $F_i(\oli{x}^*) < 0$ 
for $i \in I$ and that 
\begin{equation} \label{sum-2}
\sum_{l=0}^m  \sum_{i_1, \ldots, i_l \leq k, j_1, \ldots, j_{m-l} > k} (-1)^l r_{i_1} \ldots r_{i_l} F_{i_1} \ldots F_{i_l} \det(C_{j_1,\ldots j_{m-l}}) > 0,
\end{equation}
for each $m=0, \ldots n$, where $C_{j_1,\ldots j_{m-l}}$ is the matrix obtained by keeping rows and columns $j_1, \ldots j_{m-l}$ from $D^* A$.
\end{Thm}


\begin{proof}[Proof of the theorem]
Put 
\[
F_i(x_1,\ldots,x_n) = 1 - \frac{1}{K_i} \sum_{j=1}^n \al_{ij} x_j.
\]
We begin with the first case, where we have a feasible equilibrium. The Jacobian of the LV-system becomes
\begin{equation} \label{J}
J = \left( \begin{array}{cccc} 
r_1(F_1-\frac{x_1}{K_1}) & -\frac{r_1}{K_1} \al_{12} x_1  & \ldots & -\frac{r_1}{K_1}\al_{1n}x_1 \\
\ldots & & & \\
-\frac{r_n}{K_n} \al_{n1}x_n & -\frac{r_n}{K_n} \al_{n2}x_n & \ldots & r_n(F_n - \frac{x_n}{K_n}) 
\end{array} \right). 
\end{equation}
   If we assume that $\oli{x}^*=A^{-1} \oli{K}$ is the feasible equilibrium, then 
  $F_1(\oli{x}^*) = F_2(\oli{x}^*) = \ldots = F_n(\oli{x}^*) = 0$. Moreover, let us make a simple change of variables and let 
  \[
  y_i^* =  \frac{r_i}{K_i} x_i^* \quad \text{and} \quad  \oli{y}^* = (y_1^*, \ldots, y_n^*).
  \]
  Then, consequently, 
    \begin{align}
J &=  \left( \begin{array}{cccc} 
-y_1^* & -\al_{12}  y_1^*  & \ldots & -\al_{1n}  y_1^* \\
\ldots & & & \\
-\al_{n1}  y_n^* & -\al_{n2}  y_n^* & \ldots &  - y_n^* 
\end{array} \right) = -\oli{y}^* A. \\ 
\end{align}
  Let $J_1=\oli{y}^* A = -J$. We are looking now at the eigenvalues of this matrix. 

   With $D_{i_1,i_2, \ldots,i_k}$ we mean the principal minor obtained by deleting all the rows and columns apart from $i_1, \ldots i_k$ from $A$. (We assume that all $i_1, \ldots, i_k$ are distinct.) We have 
    \begin{align}
p(\la) = \det(J_1 - \la I) &= (-1)^n \la^n + (-1)^{n-1}\la^{n-1}(\sum_j y_j^*) + (-1)^{n-2} \la^{n-2} (\sum_{i \neq j} y_i^* y_j^* D_{ij})  \\
&+ \ldots (-1) \la ( \sum_{i_1, \ldots, i_{n-1}} 
 y_{i_1}^* \ldots y_{i_{n-1}}^*   D_{i_1,\ldots, i_{n-1}}  ) + \det(A) y_1^* \ldots y_n^*.
    \end{align}
    We have, by definition, that for each fixed $k$, 
    \[
    \sum_{i_1,\ldots,i_k} y_{i_1}^* \ldots y_{i_k}^* D_{i_1,\ldots,i_k} = \sum_{C \in \DD_k(B)} \det(C),
    \]
    which are precisely the coefficients in front of $(-\la)^{n-k}$. By assumption, it follows that there are precisely $n$ sign changes. It follows from Descartes theorem of signs that there are at most $n$ positive roots to the equation. By inspection, it is clear that there cannot be any negative roots since, for positive $x$, by letting $\la = -x < 0$, we get a polynomial with only positive terms. However, there can be complex roots. Since the polynomial is real, complex roots appear in conjugate pairs. If we factor two complex conjugate roots, we get, letting $\la = x$,
    \[
p(\la) = (x^2 + px + q)(x^{n-2} + a_{n-3}x^{n-3} + \ldots + a_1 x + a_0).
    \]
 We are assuming that the equilibrium is stable, so all eigenvalues have positive real part (hence negative if we consider $-J = \oli{y}^* (-A)$). This means that $p < 0$, $q > 0$, and $a_{n-3} < 0, a_{n-4} > 0$, etc. If we expand the above expression, we get
    \begin{equation} \label{poly}
p(\la) = x^n + (a_{n-3} + p))x^{n-1} + (a_{n-4} + pa_{n-3} + q)x^{n-2} + \ldots 
    \end{equation}
    So $p+a_{n-3} < 0$, $a_{n-4} + pa_{n-3} + q > 0$ and so on. If there are no more complex roots, then we have that $a_{k} < 0$ if $k$ is odd and vice versa. Indeed, if the equilibrium is stable, the real part of the complex roots is positive, i.e. $p$ is negative, and all real roots are positive. If there are no more complex roots, the polynomial $x^{n-2} + a_{n-3}x^{n-3} + \ldots$ has $n-2$ sign changes, and the whole polynomial (\ref{poly}) has $n$ sign changes. If there are more complex roots, factor them as second degree real polynomials as before until we only have real roots left. In the end we get
    \[
    (x^2 + p_1 x +q_1) \ldots (x^2 + p_mx +q_m) Q(x),
    \]
    where $Q$ only has real roots, where $p_j, q_j \in \R$, and where each polynomial $x^2 + p_j x + q_j$, for $1\leq j \leq m$ has two complex (non-real) roots. 
    We can then proceed in the same manner and first expand the parentheses of the $m$th second degree polynomial above and $Q$. We must have that $Q$ has $n-2m$ sign changes, and $(x^2 +p_m x + q_m)Q(x)$ has $n-2m+2$ sign changes. Continuing in the same way, 
    \[
    (x^2 + p_{m-1}x + q_{m-1})(x^2 + q_m x +q_m) Q(x)
    \]
    has $n-2m + 4$ sign changes. Hence $p$ itself has $n$ sign changes, if the equilibrium is stable. This finishes the proof in the case of a feasible equilibrium. 

Suppose now that the equilibrium point $\oli{x}^*$ is not feasible. That means that the set $I$ of indices $i$ where $x_i^* = 0$ is non-empty. Then 
\[
r_i F_i(\oli{x}^*) - \frac{r_i x_i^*}{K_i} = r_i F_i(\oli{x}^*) < 0
\]
at the equilibrium for $i \in I$ and $r_i F_i -r_i x_i^*/K_i = -(r_i/K_i) x_i^*$ at the equilibrium for $j \notin I$. 
Let us, without loss of generality, assume that $I = \{1, \ldots, k\}$ and $I^c = \{ k+1, \ldots n\}$. Let us again switch to the coordinates $y_i^* = (r_i/K_i)x_i^*$, for $i = 1, \ldots, n$. Also, put $r_i F_i = F_i^*$. Then the Jacobian assumes the form

  \begin{align}
J &= \left( \begin{array}{ccccc} 
F_1^* & 0 & \ldots & \ldots & 0 \\
0 & F_2^* & 0 & \ldots & 0 \\
\ldots & \ldots & \ldots & \ldots & \\
0 & \ldots  & 0 & F_k^* & 0 \\
-\al_{k+1, 1} y_{k+1}^* & \ldots  & -y_{k+1}^* & \ldots & -\al_{k+1,n}  y_{k+1}^* \\
\ldots & \ldots & \ldots & \ldots & \\
-\al_{n1}  y_n^* & -\al_{n2}  y_n^* & \ldots & \ldots &  - y_n^* 
\end{array} \right). \\ 
\end{align}
With $J_1 = -J$, the characteristic polynomial becomes
\begin{align}
p(\la) &= \det(J_1 - \la I) = (-1)^n \la^n + (-1)^{n-1}\la^{n-1}(- \sum_{i \leq k} F_i^* + \sum_{j > k} y_j^*) \\
&+ (-1)^{n-2} \la^{n-2} (\sum_{i, j \leq k} F_i^* F_j^* - \sum_{i \leq k, j > k} F_i^* y_j^* + \sum_{i,j > k} y_i^* y_j^* D_{ij})  \\
&+ (-1)^{n-m} \la^{n-m} \bigl( \sum_{l=0}^m  \sum_{i_1, \ldots, i_l \leq k, j_1, \ldots, j_{m-l} > k} (-1)^l F_{i_1}^* F_{i_2}^* \ldots F_{i_l}^* y_{j_1}^* \ldots y_{j_{m-l}}^* D_{j_1,\ldots j_{m-l}} \bigr) \\
&+ (-1)^k F_1^* F_2^* \ldots F_k^* D_{k+1, k+2, \ldots ,n} y_{k+1}^* \ldots y_n^*.
\end{align}
By the same argument as in the canonical case, we see that a necessary condition for stability is that all coefficients in the above polynomial are positive. The conclusion follows. 
\end{proof}

\section{Discussion}

Theorem \ref{equi-s} complements the earlier results by N. Adachi, Y. Takeuchi and H. Tokumaru and B.S. Goh, by giving necessary conditions for stability. In some sense, the necessary condition is close to saying that $A$ is a $P$-matrix.  
Since all the diagonal elements in $D^*$ are positive, it is easy to see that, if $A$ is a $P$-matrix, then the conditions of the above theorem are satisfied. Indeed, in the feasible case this follows immediately since every term in the sum (\ref{sum-1}) is positive itself for each fixed $k$ (the determinants are all positive). In the sub-feasible case, we recall that for every $j \in I$, 
\[
F_j(\oli{x}^*) < 0 \text{  for   } j \in I
\]
by, e.g., Lemma 2 in (\cite{Ad-Ta-3}) (this condition is sometimes referred to as the {\em non-invadability condition}). Hence the product $F_{i_1} \ldots F_{i_l} (-1)^l > 0$ and consequently every term in the sum (\ref{sum-2}) is positive (again using that all the determinants are positive). 

Since the necessary condition in Theorem \ref{equi-s} is a statement about a sum or a weighted sum (in the case of a sub-feasible equilibrium) of determinants of all matrices in $\DD_k(B)$,  the necessary condition seems only slightly weaker than the $P$-matrix condition. One could then ask if not the $P$-matrix condition itself is necessary for stability. The answer is in general no, and in the proof one can see that the only exception can arise if there are complex eigenvalues of the Jacobian at the equilibrium point. 




\printbibliography

\end{document}